\documentclass[twocolumn,showpacs,preprintnumbers,amsmath,amssymb,superscriptaddress]{revtex4}

\usepackage{graphicx}
\usepackage{dcolumn}
\usepackage{bm}

\newcommand{\cfrest}{\varepsilon} 
\newcommand{\DE}{W_\text{coh}} 
\newcommand{\crit}{\text{cf}} 
\newcommand{\Ncl}{N_\text{drop}} 
\newcommand{\gdotn}{\dot{\gamma}_0} 

\begin{document}

\preprint{APS/123-QED}

\title{Stability of freely falling granular streams}

\author{Stephan Ulrich}
\affiliation{Instituut-Lorentz for Theoretical Physics, Leiden, Netherlands}%


\author{Annette Zippelius}%
\affiliation{Universit\"at G\"ottingen, Institute of Theoretical Physics, Germany}%
\affiliation{
Max-Planck-Institut f\"ur Dynamik und Selbstorganisation, G\"ottingen, Germany
}%

\date{\today}

\begin{abstract}
  A freely falling stream of weakly cohesive granular particles is
  modeled and analysed with help of event driven simulations and
  continuum hydrodynamics. The
  former show a breakup of the stream into droplets, whose size is measured as
  a function of cohesive energy. Extensional flow is an exact solution
  of the  one-dimensional
  Navier- Stokes equation, corresponding to a strain rate, decaying
  like $t^{-1}$ from its initial value, $\dot{\gamma}_0$. Expanding
  around this basic state, we show that the flow is stable for short
  times, $\dot{\gamma}_0 t \ll 1$, whereas for long times, $\dot{\gamma}_0 t \gg 1$, 
  perturbations of all wavelength grow. The
  growthrate of a given wavelength depends on the instant of time when
  the fluctuation occurs, so that the observable patterns can vary
  considerably.

\end{abstract}

\maketitle


Recent experiments on granular streams have revealed many features
which are familiar from molecular liquids. Somewhat surprising was the
observation of clustering \cite{Moebius06,Royer,Scott11} in freely
falling dry granular streams which are reminiscent of the droplet
patterns observed in liquids due to surface tension. Even though tiny
attrative forces could be measured and are attributed to van der Waals
interactions or capillary bridges, the observed size of the clusters
did not agree with the predictions of Rayleigh-Plateau. In another set
of experiments~\cite{Yacine08,Prado11} capillary waves and their dispersion were measured,
allowing to deduce a (tiny) surface tension. Exciting perturbations of
a given frequency and observing their initial growth was consistent
with the Rayleigh-Plateau analysis.

In this paper, we model a freely expanding stream of weakly cohesive,
inelastically colliding grains and simulate it for the parameters
deduced from experiment. We confirm the observed clustering and
determine growth rates and drop sizes in dependence on cohesive
energy. The initial instability is analysed within a continuum
description, based on the Navier-Stokes equations. Given an exact
solution of the nonlinear equtions for extensional flow, linear
stability analysis can be performed and predicts nonmonotoneous
behaviour as a function of time: For short times a finite strainrate
stabilises the stream, wheras for long times it becomes completely unstable.

{\it Cohesive forces} --- We model~\cite{UlrichPRL} the
grains as hard spheres of diameter $d$. When two particles approach
they do not interact until they are in contact whereupon they are
inelastically reflected with a coefficient of restitution
$\cfrest$. Moving apart, the particles  feel an attractive potential of
range $d_\crit$. Such an attractive force can be due to capillary
bridges or van der Waal forces, if the particles are deformed in
collisions. As the spheres withdraw beyond the distance $d_\crit$, a
constant amount of energy $\DE$ is lost provided 
the normal relative velocity $\Delta v$ of the impacting particles is
sufficient to overcome the potential barrier, $\Delta v > \Delta
v_\text{crit} = \sqrt{2 \DE / \mu}$ (where $\mu$ is the reduced mass),
otherwise the particles form a bounded state, oscillating back and forth.

{\it Strainrate} --- We assume that the particles fall out of the
container into a vacuum \cite{Clement} with an intial velocity $v_0$. For simplicity, consider
a column of $n$ particles leaving the hopper
sequentially, with a time intervall $\Delta t = d/v_0$, and ignore
collisions for now. We notice that the $i^\text{th}$ particle will be
accelerated according to $\dot{z}_i(t) = g(t-i\Delta
t)+v_0$. 
Hence there is an \emph{initial} velocity gradient, which can be
computed from
\begin{equation}
 \frac{\Delta v}{\Delta z}=\frac{\dot{z}_{i+1}-\dot{z}_i}{z_{i+1}-z_i}= \frac{g \Delta t}{v_0\Delta t}=\frac{g}{v_0}=:\dot{\gamma}_0  \,. \label{eq:stranRateDef}
\end{equation}
In the comoving frame the stream expands freely: the particles move
with constant velocity, however their distance increases. Hence we
expect the strain rate to decrease as a function of time according to
\begin{equation}
\frac{dv}{dz}=\frac{\dot{\gamma}_0}{1+\dot{\gamma}_0t} \,. \label{eq:strainRate}
\end{equation}
This will turn out to be important for the stability analysis:
stretching is known~\cite{Eggers97} to stabilise the flow and hence prevent
clustering. As we will see below, this is precisely what happens for
short times, wheras for long times we recover the clustering
instability, when the strain rate has become sufficiently small.

{\it Simulation} --- This simple model can be simulated with an event
driven code \cite{UlrichPRL}, allowing us to consider large systems
with up to $N = 10^6$ particles. We simulate the freely falling stream
in the rest frame of the stream, imposing a homogeneous velocity
gradient $\dot{\gamma}_0 = \frac{dv}{dz} = \frac{g}{v_0}$ together
with a small random velocity in the initial state. Given the
instantaneous interactions in our simple model, the strain rate
$\gdotn$ and the cohesive energy $\DE$ are not independent parameters:
If, e.g., $\gdotn$ is increased by a factor of two and $\DE$ by a
factor of $4$, the particles follow exactly the same trajectories,
just twice as fast. Hence, we only vary the cohesive energy,
$w:=\DE/W_\text{coh,exp}$, which is convenienly measured relative to
the typical experimental value of ref.~\cite{Royer},
i.e. $W_\text{coh,exp} = 10^{-15}\,\text{J}$.  The cohesive energy,
relates to the surface tension $\Gamma$, used later, through $\Gamma
\approx \DE / d^2$ \cite{Rowlinson82,Royer}. The remaining parameters
are chosen, unless specified otherwise, to match the typical
experimental values, namely the coefficient of restitution $\cfrest =
0.9$ , stream's initial volume fraction $\phi = 0.5$, and initial
stream radius $r_0 = 19d$.

\begin{figure}[h]
 \includegraphics[width=.35\textwidth]{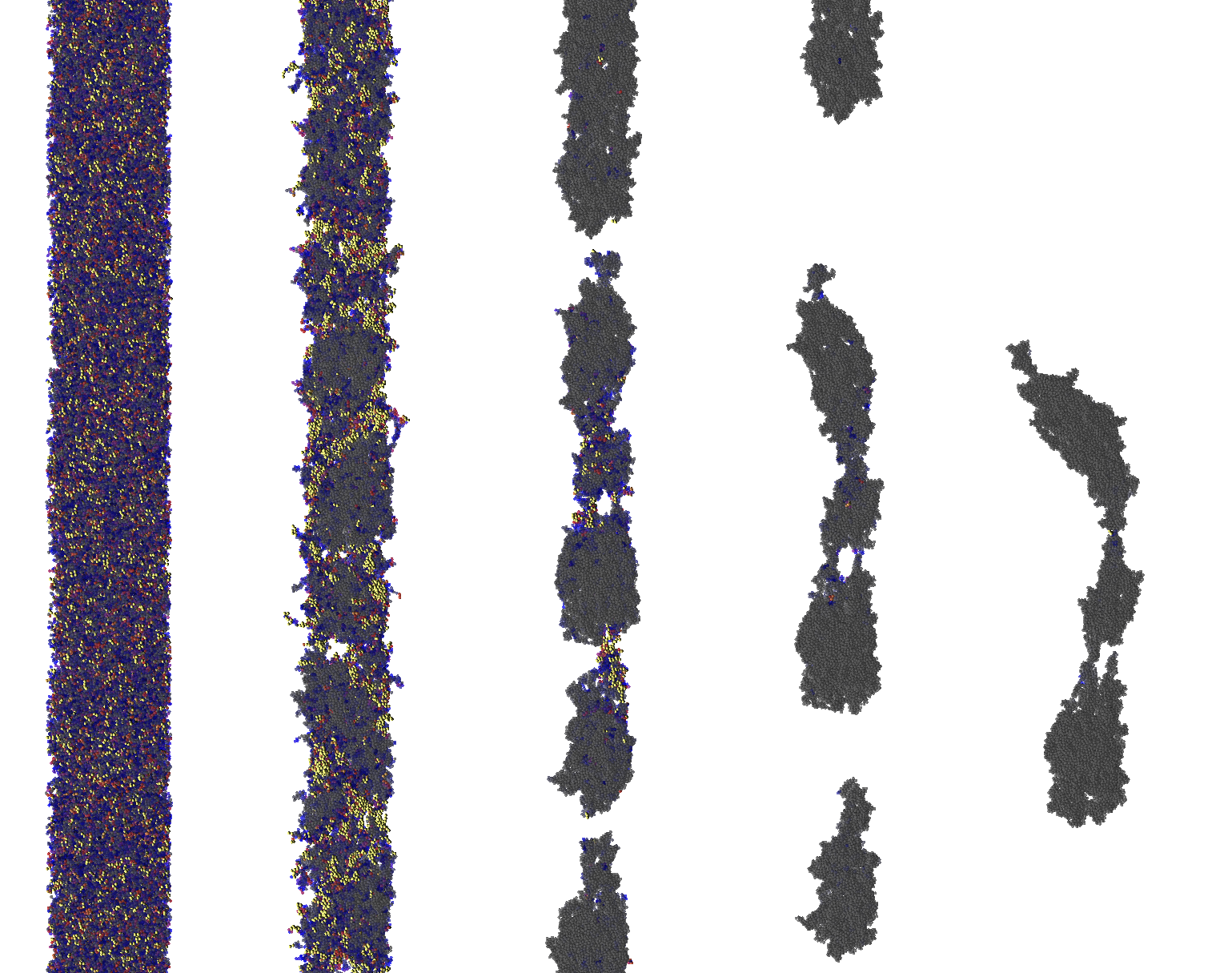}
 \centering
 \caption{(color online) Snapshots of the system for different times;
   colors indicate relative motion of adjacent particles; frozen areas
   appear in gray (dark) and yellow (bright) corresponds to areas of
   ongoing deformations; small droplets of size $\Ncl< 1000$ are
   ignored for better visibility. See \cite{movie} for a movie.}
 \label{fig:snapshots}
\end{figure}

{\it Simulation results} --- Fig.~\ref{fig:snapshots} shows snapshots of the same system at five
different times, demonstrating, how the initially straight stream
profile develops inhomogeneities which grow in time and finally lead
to separate clusters.

\begin{figure}[h]
 \includegraphics[width=.4\textwidth]{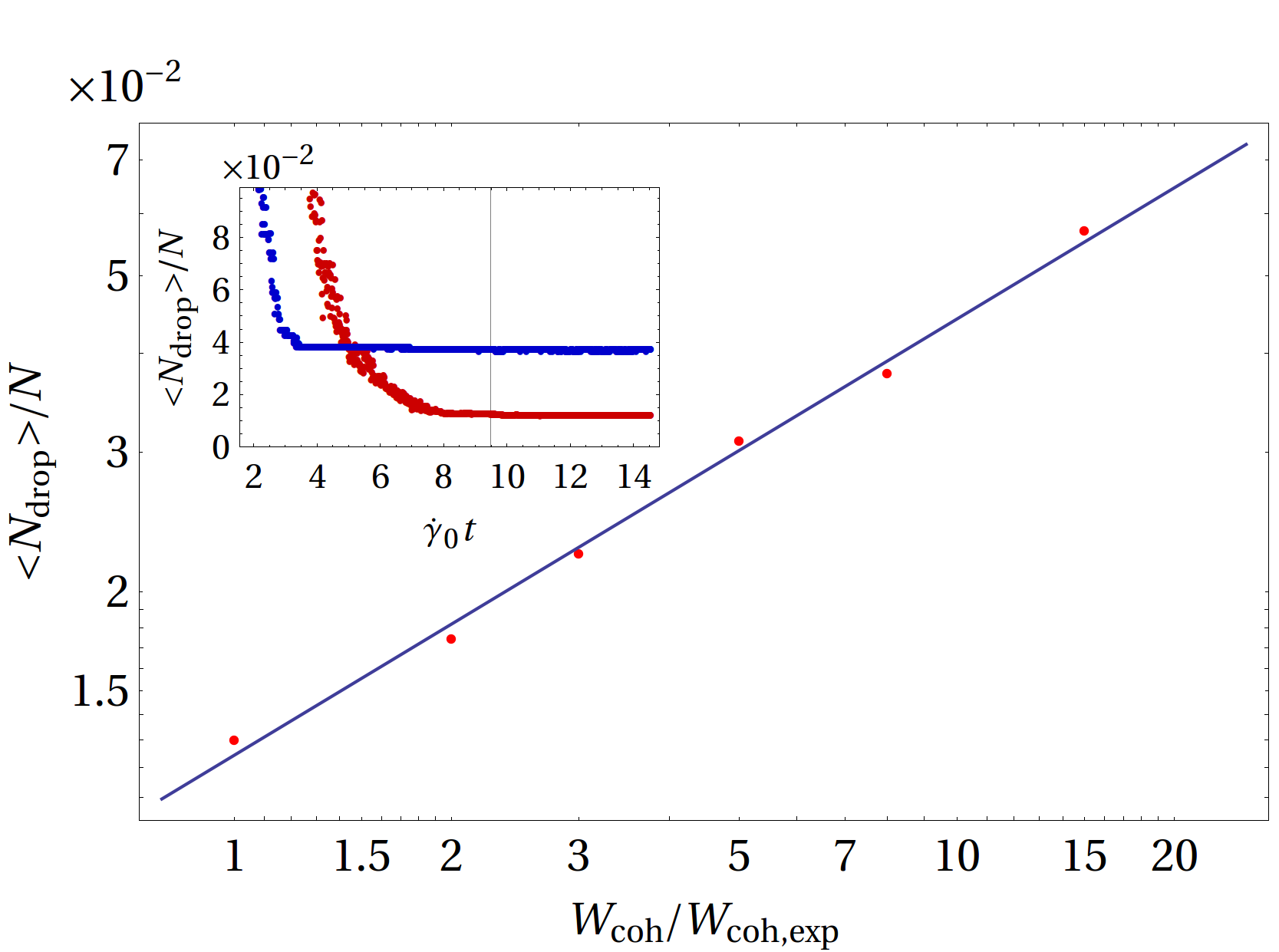}
 \centering
 \caption{(color online) Mean droplet size $\langle
   \Ncl \rangle$ as a function of $w=\DE/W_\text{coh,exp}$; data
   points are results from the simulation and the solid line is a
   power law fit; inset: mean droplet size $\langle \Ncl \rangle$
   as a function of time for $w = 8$ (high final
   value) and $w = 1$ (low final value); at the grey vertical line
   all systems have reached a steady state, which is used for the main
   plot.}
 \label{fig:NvsWcoh}
\end{figure}

In the inset of Fig.~\ref{fig:NvsWcoh} we plot the mean droplet size
$\Ncl$ as a
function of time for two values of $\DE$. After a sharp initial
decrease due separation of the stream
into clusters, $\Ncl$ reaches a steady state. Its value is shown in
the main plot for a range of cohesive energies.
Scaling arguments in \cite{Scott11} suggest that the typical length of
a droplet, rescaled back to its length on the unstretched stream,
$\lambda_0$, should scale like the square root of the cohesive
energy. Hence, we expect $\Ncl \propto \lambda_0 \propto
\DE^{1/2}$. The solid line is a fit to the data points with an
exponent $\beta = 0.54$, confirming the simple scaling arguments.

\begin{figure}[h]
 \includegraphics[width=.4\textwidth]{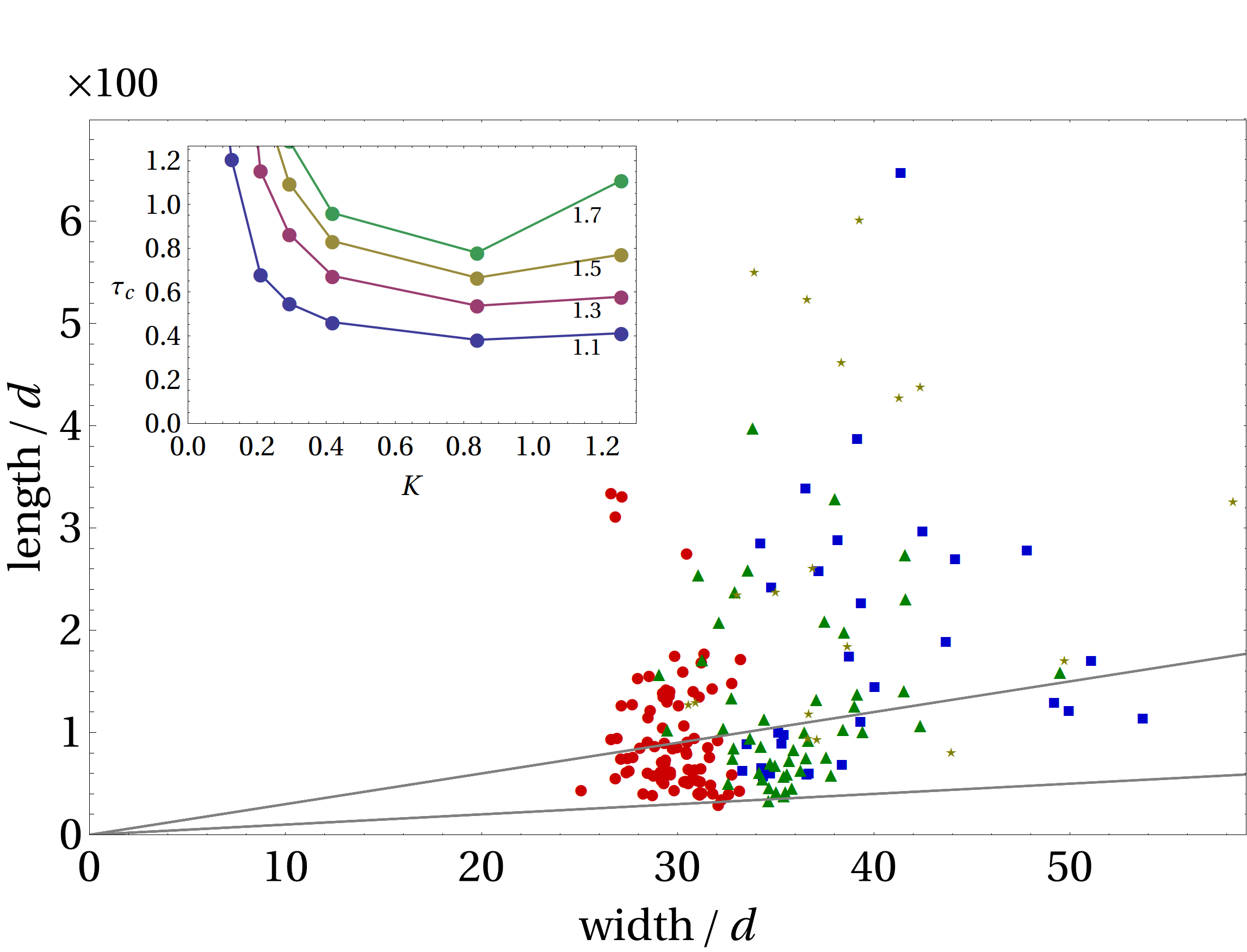}
 \centering
 \caption{(color online). Length vs.\ width of individual clusters
   with $\Ncl \ge 1000$; $w=1$ (disk), $w=3$ (triangle), $w=8$
   (square), and $w=15$ (star); gray lines correspond to aspect
   ratios of 1 and 3; inset: time $\tau_c$ for an imposed perturbation
   of wavenumber $K$ to grow beyond a given value $A_c$ (1.7, 1.5,
   1.3, 1.1 from top to bottom)}
 \label{fig:widthVsLength}
\end{figure}

The actual shape of the droplet is more difficult to capture
systematically than its mass, since it continues to change slightly
even after the droplets have separated. Royer \emph{et al.}\
\cite{Royer} characterize droplets by their length $\lambda_c$ and
width $w_c$ , right before they hit the bottom of the experimental
setup. They find that droplets' aspect ratios $\lambda_c / w_c$ always
fall in between 1 and 3. Even though the droplet formation appears to
be surface tension driven, these findings preclude the expected
Rayleigh-Plateau instability as a predominant mechanism (which only
allows aspect ratios $\ge \pi$). In Fig.~\ref{fig:widthVsLength} we
show the simulation results for the droplet lengths and width for
various $\DE$.
The most striking feature in this plot is the huge scatter in droplet
length for a given value of $\DE$. This result is at
variance with a well defined critical wavelength, corresponding to the
fastest growing mode.

Previous findings suggest that droplet formation is due to an
instability, causing small fluctuations at certain wavelengths to
grow, while other wavelengths are stable. To shed light on this
instability, we impose a small undulation $h(z) = r_0 + \epsilon
\cos(\lambda z)$ in the initial state, follow the
time evolution of the respective Fourier mode $A(\lambda,t)$
and determine the time $\tau_c$, it takes the amplitude to grow beyond a
certain value $A_c$, i.e. $A(\lambda,\tau_c)/A(\lambda,0)=A_c$.
This result is shown in the inset of
Fig.~\ref{fig:widthVsLength}. A fastest growing mode can be identified
and hardly depends on the choice of $A_c$. In the following
section, we study this instability in terms of a continuum theory and
compare the predictions to simulation and experiment.

{\it Continuum theory} --- To analyse the stability of the initially
homogeneous stream we use continuum theory
\cite{Eggers97,Eggers08}. Our starting point are the Navier-Stokes
equations for the velocity field, $\vec{v}(r,z;t)$, in cylindrical
coordinates
assuming axial symmetry
\begin{equation}
\partial_t \vec{v}+(\vec{v}\cdot \nabla)\vec{v}=\frac{\nabla
  p}{\rho}+\nu\Delta\vec{v} 
\end{equation}
together with the equation of motion for the interface $r=h(z,t)$,
\begin{equation}
\partial_t h +v_z\partial_z h=v_r|_{r=h} \,.
\end{equation}
Here $p$ denotes the pressure, $\rho$ the density and $\nu$ the shear
viscosity. These equations have to be solved, subject to the boundary
conditions, requiring the balance of normal and tangential forces at
the interface: $\sigma
\vec{n}=-\kappa\Gamma\rho\vec{n}\quad\quad$ at $r=h$.
Here $\kappa$ is the curvature of the interface, $\Gamma$ is the
surface tension divided by the density and
$\sigma_{ij}=-p\delta_{ij}+\nu (\partial_i v_j+\partial_j v_i)/\rho$
denotes the stress tensor.

To obtain approximate solutions to the above equations, we
follow Eggers \cite{Eggers97} and assume that variations in the radial
direction take place on scales small compared to variations along the
stream. Under these assumtions a one
dimensional Navier Stokes equation for $v=v_z(z,t)$ has been derived
\cite{Eggers97} for an incompressible fluid:
\begin{eqnarray}
\dot{v}+v v'&=&-\gamma\frac{\kappa'}{\rho}+3\nu\frac{(v'h^2)'}{h^2}
\\
\dot{h^2}+(v h^2)'&=&0
\end{eqnarray}

These equations have been studied in various circumstances for
molecular fluids \cite{Eggers97,Eggers08}. The best known one is the
Rayleigh Plateau instability, where one expands around a state with
constant radius and velocity which does not apply in the presence of
gravity. Jet flow dominated by viscous effects \cite{Sauter,Bohr} has
also been analysed within the above one-dimensional model. Here we
consider instead a freely falling stream \cite{Frankel} in the
comoving frame. This state is characterized by a time dependent
velocity gradient, that is constant in space:
$\overline{v}(z,t)=\frac{z \dot{\gamma}_0}{1+\dot{\gamma}_0t}$.
Incompressibilty requires
$\overline{h}(z,t)=r_0(1+\dot{\gamma}_0t)^{-1/2}$. These fields solve
the above equations {\it exactly}, allowing us to do a linear
stability analysis by expanding around the above solution.

We introduce a dimensionless position variable $Z := z/(r_0(1 +
\dot\gamma_0 t))$ such that $Z$ remains fixed, if $z$ moves along with
the stream.  The $Z$-dependence can then be taken care of by plane
waves: $\sim \exp{(ikr_0Z)}$. To further simplify the notation, we
introduce dimensionless time $\tau=\gamma_0 t$ and wavenumber $K=k r_0
$.  We obtain two linear equations for $h(z,t)-\overline{h}(z,t)
=\exp{(iKZ)}\epsilon_R(\dot{\gamma}_0t)$ and
$v(z,t)-\overline{v}(z,t)=\exp{(iKZ)}
\dot{\gamma}_0\epsilon_V(\dot{\gamma}_0t)$:
\begin{eqnarray}
  \dot {\epsilon}_V(\tau)&=&-\frac{\epsilon_V(\tau)}{1+\tau}
+ \,iK\left(\tilde{\Gamma}
-\frac{\tilde{\Gamma}K^2}{(1+\tau)^3}\right)\epsilon_R(\tau)\nonumber\\
  \dot{\epsilon}_R(\tau)&=&-\frac{\epsilon_R(\tau)}{2(1+\tau)}-
  \frac{iK\epsilon_V(\tau)}{2(1+\tau)^{3/2}} \label{eq:epsODE}
\end{eqnarray}
For clarity of presentation we have set $\nu=0$ and hence are left with one
dimensionless parameter $\tilde{\Gamma}=\Gamma/(r_0^3\gamma_0^2)$.
The generalisation to finite viscosity is straighforward and given in
the supplementary material \cite{supplementCalc}.

The above eqs.\ are two ordinary differential equations with {\it
  time-dependent} coefficients. This makes the stability analysis
complex, because a given wavenumber, which is stable at $t_0$, can
override an initially unstable mode in the course of time. Of course
the equations can easily be integrated numerically. Before discussing
the generalised eigenvalue problem, we try to extract the qualitative
behaviour by inspecting the equations for small and large times. For
$\tau=\dot{\gamma}_0 t\lesssim 1$ we expect the initial strainrate to have
a stabilising effect, in particular for long wavelength perturbations.
For long times, $\tau=\dot{\gamma}_0 t\gtrsim 1$, on the other hand, the
strainrate decays. With the ansatz ansatz $\epsilon_V,\epsilon_H\sim
e^{\lambda \tau}$ one finds for $\tau\gg 1$
\begin{equation}
\lambda_{\pm}=-\frac{3}{4(1+\tau)} \pm \left(\frac{\tilde{\Gamma}
    K^2}{2(1+\tau)^{3/2}}+\frac{1}{16(1+\tau)^2}\right)^{1/2}\nonumber
\end{equation} 
that {\it all wavenumbers are unstable} for sufficiently large $\tau$,
because the dominant term is the one involving the surface tension
($\tilde{\Gamma}$).
\begin{figure}[h]
 \includegraphics[width=.49\textwidth]{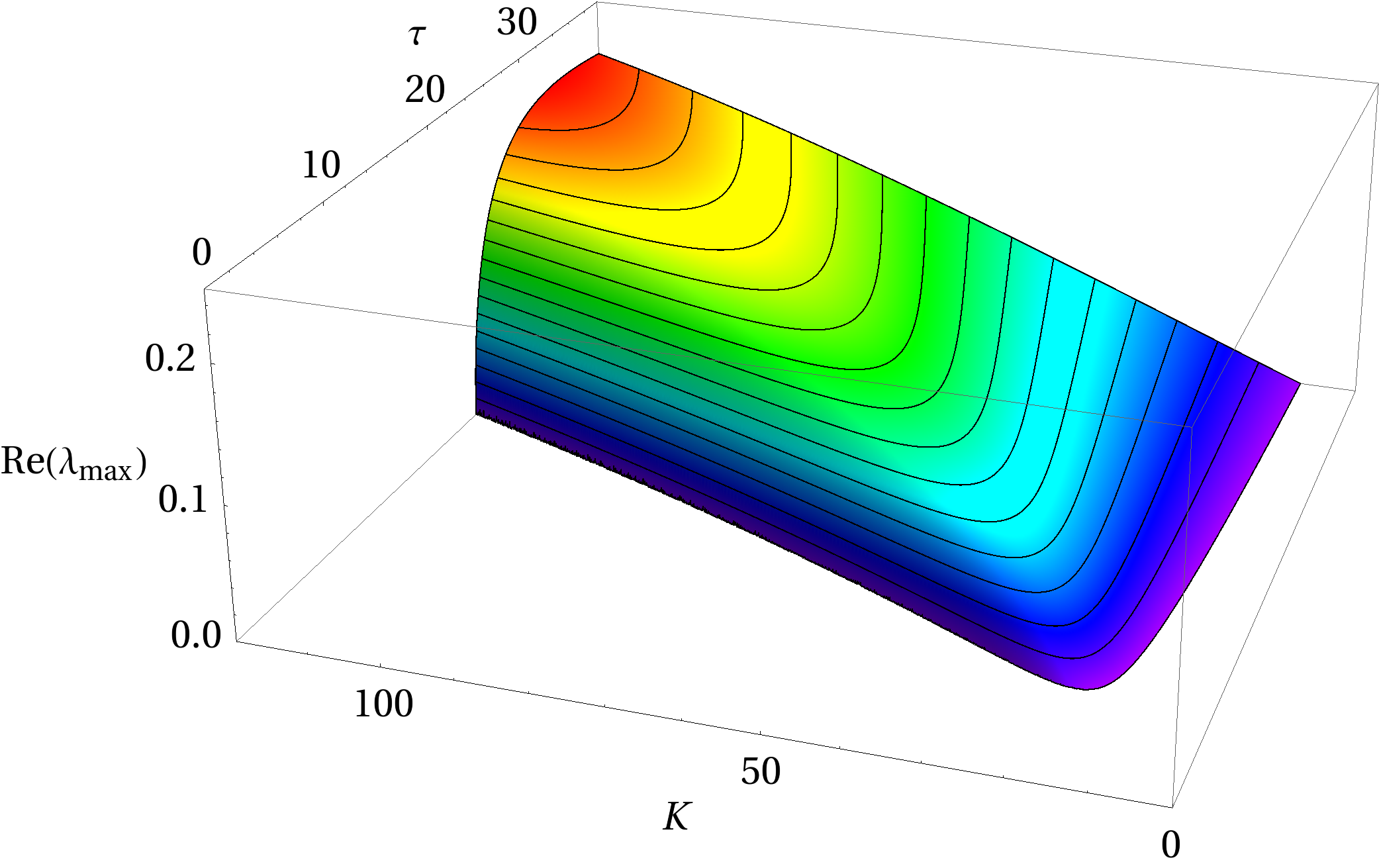}
 \centering
 \caption{Real part of the unstable eigenvalue as a function of
   wavenumber and time} 
 \label{fig:eigenvalue}
\end{figure}
To discuss the general case, we compute the eigenvalues for all $K$
and $t$, by diagonalising the time-dependent matrix of
coefficients. The larger eigenvalue, $\lambda_\text{max}$, -- responsible
for the instability -- is shown in Fig.~\ref{fig:eigenvalue} as a
function of $K$ and $t$ in the range of values, where
$\lambda_\text{max}>0$. The parameters for the initial strain
rate, $\gamma_0$, and the surface tension, $\Gamma$, are taken from
experiment and the viscosity is set to zero. We observe that initially
all wavelength are stable for $t=0$, the first instability sets in at
$\tau \approx 8$ and $K \approx 15$. This wavenumber is the initially imposed
one and has to be scaled down by the stretching factor $1+\tau$, when
the stream has been stretched up to time $\tau$. Furthermore,
wavenumbers are measured in units of the initial radius of the stream,
which decreases according to $r(\tau)=r_0/(1+\tau)^{1/2}$. Hence to
obtain the ratio of wavelength to radius at time $\tau$, when the
instability occurs, we have to scale wavenumbers according to $\tilde
K(\tau)=K (1+\tau)^{-3/2}$. If one plots the eigenvalue versus
$\tilde K(\tau)$, one observes a ridge at
approximately $\tilde{K} \approx 1$, implying that at each instant of time
$\tau$ -- while the stream is stretching -- unstable modes with roughly
the wavelength of Rayleigh Plateau are growing.
\begin{figure}[h]
 \includegraphics[width=.49\textwidth]{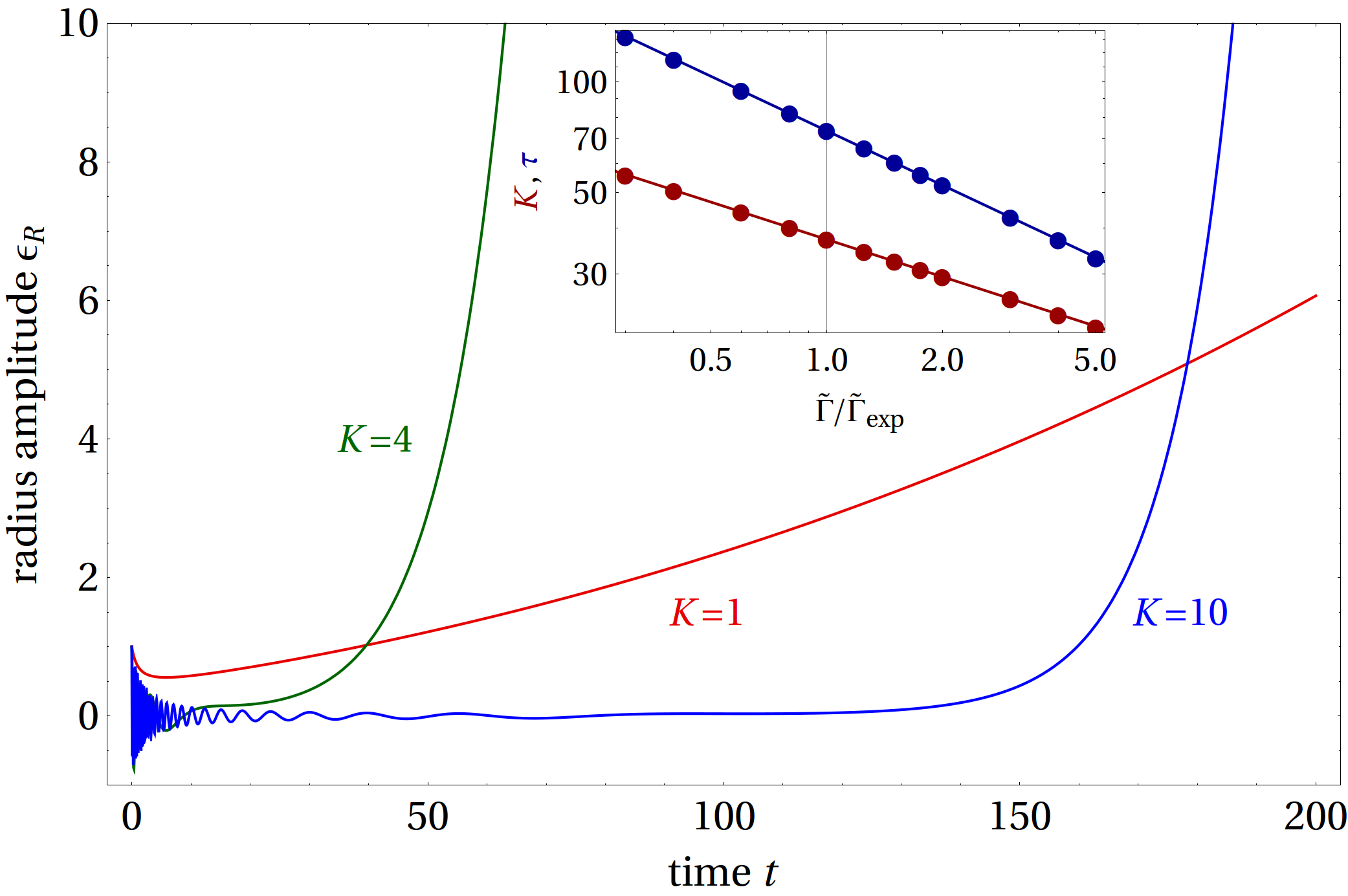}
 \centering
 \caption{Example for growth of perturbations, obtained from
   integrating the linearised equations; inset: most unstable
   wavenumber and time of growth to increase by $50$\% in dependence
   on $\Gamma/\Gamma_\text{exp}$.} 
 \label{fig:integration}
\end{figure}

However, we stress that our system is not in a stationary state, but
expanding. As a consequence the eigenvalues are time dependent, and
whether a perturbation increases or decreases depends on the time of
its occurence. In Fig.~\ref{fig:integration} we show an example of the
integrated Eqs.\ with 
3 modes excited initially ($\epsilon_R=1$ for $K=1,4,10$). As expected
from the eigenvalue analysis, initially all modes are stable, then the
smallest $K$ starts to grow, but is - at later times - overridden by
the larger wavenumbers. This is meant to illustrate that the observed
pattern will depend on the instant, when a fluctuation with a
particular wavenumber occurs.

{\it Variation of the parameters} --- 
If the viscosity is increased, the results remain qualitatively the same,
but the instabilities occur at later times, because viscosity tends to
stabilise the flow. If the initial strain rate is put to
zero, we recover the Rayleigh-Plateau instability. 
Interesting effects are observed by varying the cohesive
energy. Decreasing the cohesive energy, $\Gamma<\Gamma_\text{exp}$, the initial
range of unstable wavenumbers shifts to larger $K$, i.e. smaller
wavelength in agreement with simulations (see
Fig.~\ref{fig:NvsWcoh}). We determine the time $T_{\alpha}(K)$ which it
takes an unstable mode of wavenumber $K$ to grow by 50\% from its
initial value for several values of cohesive energy. In that way we
can deduce the critical wavenumber and the time for the stability to
occur as a function of $\Gamma$. These are shown in the inset of
Fig.~\ref{fig:integration}. We clearly observe an increase in
critical wavelength with $\Gamma$ in agreement with simulation and
experiment. Furthermore for increased $\Gamma$ the instability occurs
earlier.

{\it Conclusions} --- We have shown that a stream of granular particles
falling under gravity is generically unstable due to surface tension
-- even though the Rayleigh-Plateau argument does not apply. In the
comoving frame the stream is freely expanding, implying that the
initially straight profile is subject to a time-dependent strain rate.
Linearising the Navier-Stokes equation around this nonstationary
state, we have shown that the strain rate stabilises the straight flow
profile at short times, whereas for long times all wavenumbers are
unstable. Since we expand around a nonstationary state, the growth
rate of a given wavelength depends on the time, when the corresponding
fluctuation occurs spontaneosuly or is introduced into the flow.
Thus a variety of patterns may be observed including behaviour
reminiscent of the Rayleigh-Plateau instability~\cite{Prado11}.\\
\indent{\it Acknowledgments} -- We are grateful to Heinrich Jaeger for suggesting to apply a model
of cohesive particles to freely falling streams. Furthermore we
acknowledge interesting discussions with him and Scott Waitukaitis
as well as financial support by the Deutsche Forschungsgemeinschaft
(DFG) through Grant Zi 209/8-01.

\end{document}